\documentclass[twocolumn,showpacs,preprintnumbers,amsmath,amssymb,floatfix,a4paper]{revtex4}

\usepackage{graphicx}
\usepackage{dcolumn}
\usepackage{bm}
\usepackage{setspace}
\usepackage{geometry}
\usepackage{amsmath}

\begin{document}

\title{Investigation of thermonuclear $^{18}$Ne($\alpha$,$p$)$^{21}$Na rate via resonant elastic scattering of $^{21}$Na+$p$}

\author{L.Y.~Zhang$^{1}$}
\author{J.J.~He$^1$}
\email{jianjunhe@impcas.ac.cn}
\author{A.~Parikh$^{2,3}$}
\author{S.W.~Xu$^{1,4}$}
\author{H.~Yamaguchi$^5$}
\author{D.~Kahl$^{5}$}
\author{S.~Kubono$^{1,5,6}$}
\author{P.~Mohr$^{7,8}$}
\author{J.~Hu$^{1}$}
\author{P.~Ma$^{1}$}
\author{S.Z.~Chen$^{1,4}$}
\author{Y.~Wakabayashi$^{9}$}
\altaffiliation[Present address: ]{RIKEN (The Institute of Physical and Chemical Research), Wako, Saitama 351-0198, Japan.}
\author{H.W.~Wang$^{10}$}
\author{W.D.~Tian$^{10}$}
\author{R.F.~Chen$^{1}$}
\author{B.~Guo$^{11}$}
\author{T.~Hashimoto$^{5}$}
\altaffiliation[Present address: ]{RCNP, Osaka University, 10-1 Mihogaoka, Ibaraki, Osaka, 567-0047, Japan.}
\author{Y.~Togano$^{6}$}
\author{S.~Hayakawa$^{5}$}
\author{T.~Teranishi$^{12}$}
\author{N.~Iwasa$^{13}$}
\author{T.~Yamada$^{13}$}
\author{T.~Komatsubara$^{14}$}
\author{Y.H.~Zhang$^{1}$}
\author{X.H.~Zhou$^{1}$}

\affiliation{$^1$Institute of Modern Physics, Chinese Academy of Sciences, Lanzhou 730000, China}
\affiliation{$^2$Departament de F\'{\i}sica i Enginyeria Nuclear, EUETIB, Universitat Polit\`{e}cnica de Catalunya, Barcelona E-08036, Spain}
\affiliation{$^3$Institut d'Estudis Espacials de Catalunya, Barcelona E-08034, Spain}
\affiliation{$^4$University of Chinese Academy of Sciences, Beijing 100049, China}
\affiliation{$^5$Center for Nuclear Study, The University of Tokyo, RIKEN campus, Wako, Saitama 351-0198, Japan}
\affiliation{$^6$RIKEN (The Institute of Physical and Chemical Research), Wako, Saitama 351-0198, Japan}
\affiliation{$^7$Diakonie-Klinikum, Schw\"{a}bisch Hall D-74523, Germany}
\affiliation{$^8$Institute of Nuclear Research (ATOMKI), Debrecen H-4001, Hungary}
\affiliation{$^9$Advanced Science Research Center, Japan Atomic Energy Agency, Ibaraki 319-1106, Japan}
\affiliation{$^{10}$Shanghai Institute of Applied Physics, Chinese Academy of Sciences, Shanghai 201800, China}
\affiliation{$^{11}$China Institute of Atomic Energy, P.O. Box 275(46), Beijing 102413, China}
\affiliation{$^{12}$Department of Physics, Kyushu University, 6-10-1 Hakozaki, Fukuoka 812-8581, Japan}
\affiliation{$^{13}$Department of Physics, University of Tohoku, Miyagi 980-8578, Japan}
\affiliation{$^{14}$Department of Physics, University of Tsukuba, Ibaraki 305-8571, Japan}

\date{\today}

\begin{abstract}
The $^{18}$Ne($\alpha$,$p$)$^{21}$Na reaction is thought to be one of the key breakout reactions from the hot CNO cycles
to the rp-process in type I x-ray bursts. In this work, the resonant properties of the compound nucleus $^{22}$Mg have been
investigated by measuring the resonant elastic scattering of $^{21}$Na+$p$. An 89 MeV $^{21}$Na radioactive beam delivered
from the CNS Radioactive Ion Beam Separator bombarded an 8.8 mg/cm$^2$ thick polyethylene (CH$_{2}$)$_{n}$ target. The
$^{21}$Na beam intensity was about 2$\times$10$^{5}$ pps, with a purity of about 70\% on target. The recoiled protons
were measured at the center-of-mass scattering angles of $\theta_{c.m.}$$\approx$175.2${^\circ}$, 152.2${^\circ}$, and
150.5${^\circ}$ by three sets of $\Delta E$-$E$ telescopes, respectively. The excitation function was obtained with the
thick-target method over energies $E_x$($^{22}$Mg)=5.5--9.2 MeV. In total, 23 states above the proton-threshold in $^{22}$Mg
were observed, and their resonant parameters were determined via an $R$-matrix analysis of the excitation functions.
We have made several new $J^{\pi}$ assignments and confirmed some tentative assignments made in previous work.
The thermonuclear $^{18}$Ne($\alpha$,$p$)$^{21}$Na rate has been recalculated based on our recommended spin-parity assignments.
The astrophysical impact of our new rate has been investigated through one-zone postprocessing x-ray burst calculations.
We find that the $^{18}$Ne($\alpha$,$p$)$^{21}$Na rate significantly affects the peak nuclear energy generation rate, reaction
fluxes, as well as the onset temperature of this breakout reaction in these astrophysical phenomena.
\end{abstract}

\pacs{25.60.-t, 23.50.+z, 26.50.+x,  27.30.+t}

`\maketitle

\section{Introduction}
\label{sec:introduction}

Explosive hydrogen and helium burning are thought to be the main source of energy generation and nuclear trajectory to higher mass on the proton-rich
side of the nuclear chart in type I x-ray bursts (XRBs)~\cite{champage,wiescher,bib:lew93,bib:str06,bib:par13}.
XRBs are characterized by a sudden increase of x-ray emission within only a few seconds to a total energy output of about 10$^{40}$ ergs, which is
observed to repeat with some regularity. The recurrence time for single bursts can range from hours to days at the
typical temperature of 0.4--2 GK. The bursts have been interpreted as being generated by thermonuclear runaway on the surface of a neutron star that
accretes H- and He-rich material from a less evolved companion star in a close binary system. In XRBs, the hydrogen burning initially occurs via the
hot CNO cycle (HCNO):
\begin{center}
\begin{spacing}{2.0}
$^{12}$C($p$,$\gamma$)$^{13}$N($p$,$\gamma$)$^{14}$O($e^{+}\nu$)$^{14}$N($p$,$\gamma$)$^{15}$O($e^{+}\nu$)
$^{15}$N($p$,$\alpha$)$^{12}$C,
\end{spacing}
\end{center}
while the $^{13}$N($e^{+}\nu$)$^{13}$C $\beta$-decay in the CNO cycle is bypassed by the $^{13}$N($p$,$\gamma$)$^{14}$O reaction.
The temperature of the accretion envelope increases as the compressing and exothermic nuclear reactions going on. When the temperature
reaches about 0.4 GK, the second HCNO cycle becomes dominant:
\begin{center}
\begin{spacing}{2.0}
$^{12}$C($p$,$\gamma$)$^{13}$N($p$,$\gamma$)$^{14}$O($\alpha$,$p$)$^{17}$F($p$,$\gamma$)$^{18}$Ne($e^{+}\nu$)$^{18}$F($p$,$\alpha$)
$^{15}$O($e^{+}\nu$)$^{15}$N($p$,$\alpha$)$^{12}$C.
\end{spacing}
\end{center}
It was predicted~\cite{champage,wiescher} that the $^{18}$Ne waiting point nucleus in the second HCNO cycle could be bypassed by the
$^{18}$Ne($\alpha$,$p$)$^{21}$Na reaction at $T$$\approx$0.6 GK, and subsequently, the reaction chain breaks out, eventually leading to the
rp-process~\cite{wal81,sch98,woo04}. However, over stellar temperatures achieved in XRBs, this rate has not been sufficiently well determined.

The thermonuclear $^{18}$Ne($\alpha$,$p$)$^{21}$Na rate is thought to be dominated by contributions from resonances in the compound
nucleus $^{22}$Mg above the $\alpha$ threshold at $Q_\alpha$=8.142 MeV~\cite{wang}. As for XRBs, the temperature region of interest is
about 0.4--2.0 GK, corresponding to an excitation region of $E_x$=8.6--11.0 MeV in $^{22}$Mg. G\"{o}rres {\it et al.} made the first
estimate~\cite{gorres} of this rate with rather limited experimental level-structure information in $^{22}$Mg. The energies for the
$^{22}$Mg resonances were estimated simply by shifting those of known natural-parity states in the mirror $^{22}$Ne by a fixed amount
(about 200 keV). The uncertainty of this first rate was mainly caused by the errors in resonant energies (or excitation energies) and
resonant strengths of the excited states above the $\alpha$ threshold in $^{22}$Mg. After that, the precise locations of the excited states
in $^{22}$Mg were studied extensively by many transfer reaction experiments. For example, Chen {\it et al.}~\cite{chen} determined the
excitation energies with a typical uncertainty of 20--30 keV in a $^{12}$C($^{16}$O,$^{6}$He)$^{22}$Mg experiment. However, the spin-parity
assignments assumed and the spectroscopic $S_\alpha$ factors adopted following the idea of G\"{o}rres {\it et al.} were still uncertain.
Caggiano {\it et al.}~\cite{caggiano} and Berg {\it et al.}~\cite{berg} measured the excitation energies with a better precision
(about 10--20 keV), but no spin-parity assignment was given. Later, Matic {\it et al.}~\cite{matic} measured the excitation energies
precisely by a $^{24}$Mg($p$,$t$)$^{22}$Mg experiment, with uncertainty of about 1--15 keV achieved for most states above the $\alpha$
threshold; the spin-parity assignments were tentatively made based on the shell-model calculation or those of mirror states in $^{22}$Ne.
Thus, the thermonuclear $^{18}$Ne($\alpha$,$p$)$^{21}$Na rate was constrained very well in the resonant energy aspect.
In a later $^{24}$Mg($p$,$t$)$^{22}$Mg measurement, Chae {\it et al.}~\cite{chae} observed six excited states in $^{22}$Mg above the $\alpha$
threshold, and some spin-parity assignments were made via an angular distribution measurement. However, the insufficient resolution of
their measurement at the center-of-mass (c.m.) scattering angles of $\theta_{c.m.}$ above 20${^\circ}$ made such $J^{\pi}$ assignments
questionable~\cite{matic} ({\it e.g.}, the 8.495 MeV peak was contaminated by the nearby 8.572 and 8.658 MeV states as shown in their
Fig. 3). In our previous experiment of $^{21}$Na+$p$ resonant elastic scattering~\cite{he_epja,he}, new spin-parity assignments were made
only tentatively for the the 8.547 and 8.614 MeV states in $^{22}$Mg due to low statistics. Those assignments gave a quite different rate
for the $^{18}$Ne($\alpha$,$p$)$^{21}$Na reaction compared to the rate estimated in Chen {\it et al.} work~\cite{chen}. Such tentative
assignments clearly motivate further investigation.

A comparison of all available reaction rates of $^{18}$Ne($\alpha$,$p$)$^{21}$Na shows discrepancies of up to several orders of magnitude
around $T$$\sim$1 GK~\cite{matic}. So far, more than 40 levels (up to $E_x$=13.01 MeV) have been observed above the $\alpha$ threshold in
$^{22}$Mg. Such high level density suggests that a statistical model approach might provide a reliable estimate of the rate. However, only
natural-parity states in $^{22}$Mg can be populated by the $^{18}$Ne+$\alpha$ channel, and thus the effective level density will be considerably
lower. It remains unclear whether the statistical-model calculations provide a reliable rate estimation in a wide temperature region~\cite{matic}.
There are still many resonances (above the $\alpha$ threshold) without firm spin-parity assignments, which need to be determined experimentally.
As a consequence, the accuracy of the current $^{18}$Ne($\alpha$,$p$)$^{21}$Na reaction rate is mainly limited by the lack of experimental
spin-parities and $\alpha$ partial widths $\Gamma_\alpha$ (or spectroscopic factors $S_\alpha$) of the resonances in $^{22}$Mg above the $\alpha$ threshold.

So far, only two direct measurements~\cite{bradfield,groombridge} for the $^{18}$Ne($\alpha$,$p$)$^{21}$Na reaction were reported. The lowest
energies achieved in these studies ($E_{c.m.}$=2.0 and 1.7 MeV) are still too high compared with the energy region $E_{c.m.}$$\leq $1.5 MeV of
interest for HCNO breakout in XRBs. New results~\cite{salter} have recently become available, which determined the $^{18}$Ne($\alpha$,$p_0$)$^{21}$Na
cross sections in the energy region of $E_{c.m.}$=1.19--2.57 MeV by measuring those of the time-reversal reaction
$^{21}$Na($p$,$\alpha$)$^{18}$Ne in inverse kinematics. In addition, similar experiments were performed at the Argonne National Laboratory (ANL),
but the results were only reported in the ANL annual reports~\cite{anl}. The ANL cross section data are consistent with those in Ref.~\cite{salter}.
Nonetheless, these results are still insufficient for a reliable rate calculation at all temperatures encountered within XRBs. Recently, a new
reaction rate was recommended based on the combined analysis of all literature data~\cite{matic2}.

In this work, the $^{18}$Ne($\alpha$,$p$)$^{21}$Na rate is determined via the measurement of the resonant elastic scattering of
$^{21}$Na+$p$. This is an entirely new high-statistics experiment compared to the previous one~\cite{he_epja,he}. In the resonant
elastic-scattering mechanism, $^{22}$Mg is formed via the fusion of $^{21}$Na+$p$ as an excited compound nucleus, whose
states promptly decay back into $^{21}$Na+$p$. This process interferes with Coulomb scattering resulting in a characteristic resonance
pattern in the excitation function~\cite{ruiz}. With this approach, the excitation function was obtained simultaneously in a wide range of
5.5--9.2 MeV in $^{22}$Mg with a well-established thick-target method~\cite{art90,gal91,kub01}. In total, 23 states above the proton-threshold
in $^{22}$Mg were observed, and their resonant parameters were determined via an $R$-matrix analysis of the experimental data.
Part of the experimental results previously reported in Ref.~\cite{he_prc_rc} is revisited through a more detailed analysis. The detailed
experimental results presented here supersede those of Ref.~\cite{he_prc_rc}.

\section{Experiment}
\label{sec:experiment}
\begin{figure}
\resizebox{8.3cm}{!}{\includegraphics{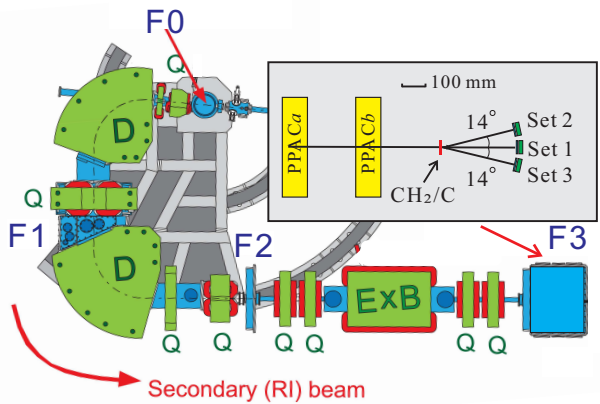}}
\caption{\label{fig1_crib} (Color online) Schematic view of CRIB utilized for the $^{21}$Na beam production. The experimental setup for
measurement of the $^{21}$Na+$p$ scattering was installed in the chamber at F3.}
\end{figure}

The experiment was carried out at the CNS Radioactive Ion Beam separator (CRIB)~\cite{kub02,yan05}, installed by the Center for Nuclear Study
(CNS), University of Tokyo in the RIKEN Nishina Center. During the last decade, the radioactive ion beams (RIBs) produced at
CRIB have been successfully utilized in the resonant scattering experiments with a thick-target method~\cite{ter03,ter07,he_al,yam09,he},
which proved to be a successful technique as adopted in the present study. A schematic view of CRIB and the measurement setup are shown in
Fig.~\ref{fig1_crib}. An 8.2 MeV/nucleon primary beam of $^{20}$Ne$^{8+}$ was accelerated by an AVF cyclotron ($K$=79) at RIKEN, with an
average intensity of 65 pnA. At the primary focal plane (F0), the beam bombarded a liquid nitrogen-cooled $D_{2}$ gas target (90 K)~\cite{yamaguchi}.
The gas was confined in a cylindrical chamber (length=80 mm, $\phi$=20 mm) whose entrance and exit windows were each made of 2.5~$\mu$m thick
Havar foils. The effective thickness of $D_{2}$ gas was about 2.86~mg/cm$^{2}$ at a pressure of about 530~Torr. The $^{21}$Na beam was
produced via the $^{20}$Ne($d$,$n$)$^{21}$Na reaction in inverse kinematics, and separated subsequently by two dipoles and a Wien Filter.
At the momentum dispersive focal plane (F1), a slit of $\pm$5 mm was installed to remove the contamination from the secondary beam. After F1, the dipole selected the $^{21}$Na$^{11+}$ particles at a mean energy of 5.9 MeV/nucleon with a momentum spread of
$\pm$0.3\%. The Wien Filter was operated at a high-voltage of $\pm$70 kV. Finally, a purity of about 70\% was achieved for the $^{21}$Na beam on the
secondary target.

\begin{figure}[!h]
\resizebox{8.3cm}{!}{\includegraphics{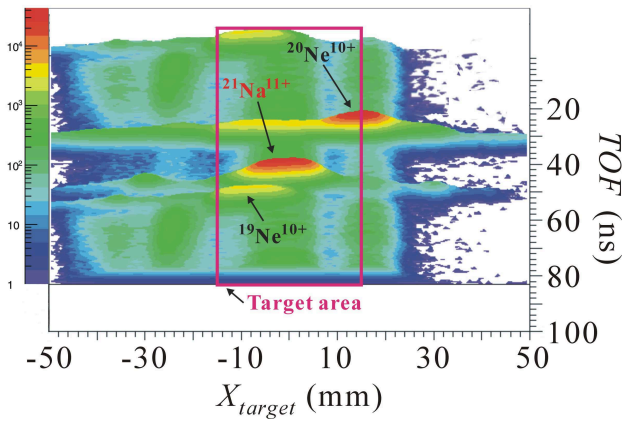}}
\caption{\label{fig2_beam} (Color online) 3D contour plot for the beam particles identification. See text for details.}
\end{figure}

The setup in a scattering chamber at the experimental focal plane (F3) consisted of two parallel-plate avalanche counters (PPACs)~\cite{kum01},
an 8.8 mg/cm$^2$ thick polyethylene (CH$_{2})_{n}$ target, and three sets of $\Delta$E-E silicon telescopes. The PPACs measured the timing and
two-dimensional position information of the incoming beam, and determined the beam position on the secondary target during the measurement.
the beam identification plot is shown in Fig.~\ref{fig2_beam}. It shows that the beam particles were clearly identified in an event-by-event mode.
Here, $TOF$ is the time-of-flight between PPACa and the RF signal from the cyclotron, which is equivalent to the beam flight time from F0 to F3.
The beam position on the target ({\it i.e.}, $X_{target}$ in Fig.~\ref{fig2_beam}) was determined by the hitting positions on the two PPACs.
The beam impinged on an 8.8 mg/cm$^2$ polyethylene (CH$_2$)$_n$ target, which was thick enough to stop all the beam ions. Here, the energy of
$^{21}$Na beam was 89.4 MeV with spread of 1.95 MeV (FWHM) on the target. The beam-spot widths (FWHM) were 9.5 mm in horizontal and 4.8 mm in
vertical directions. The horizontal and vertical angular spread (FWHM) were 10 mrad and 22 mrad, respectively. The averaged intensity of
$^{21}$Na beam was about 2$\times$10$^{5}$ pps on the target.

\begin{figure}[!h]
\resizebox{8.3cm}{!}{\includegraphics{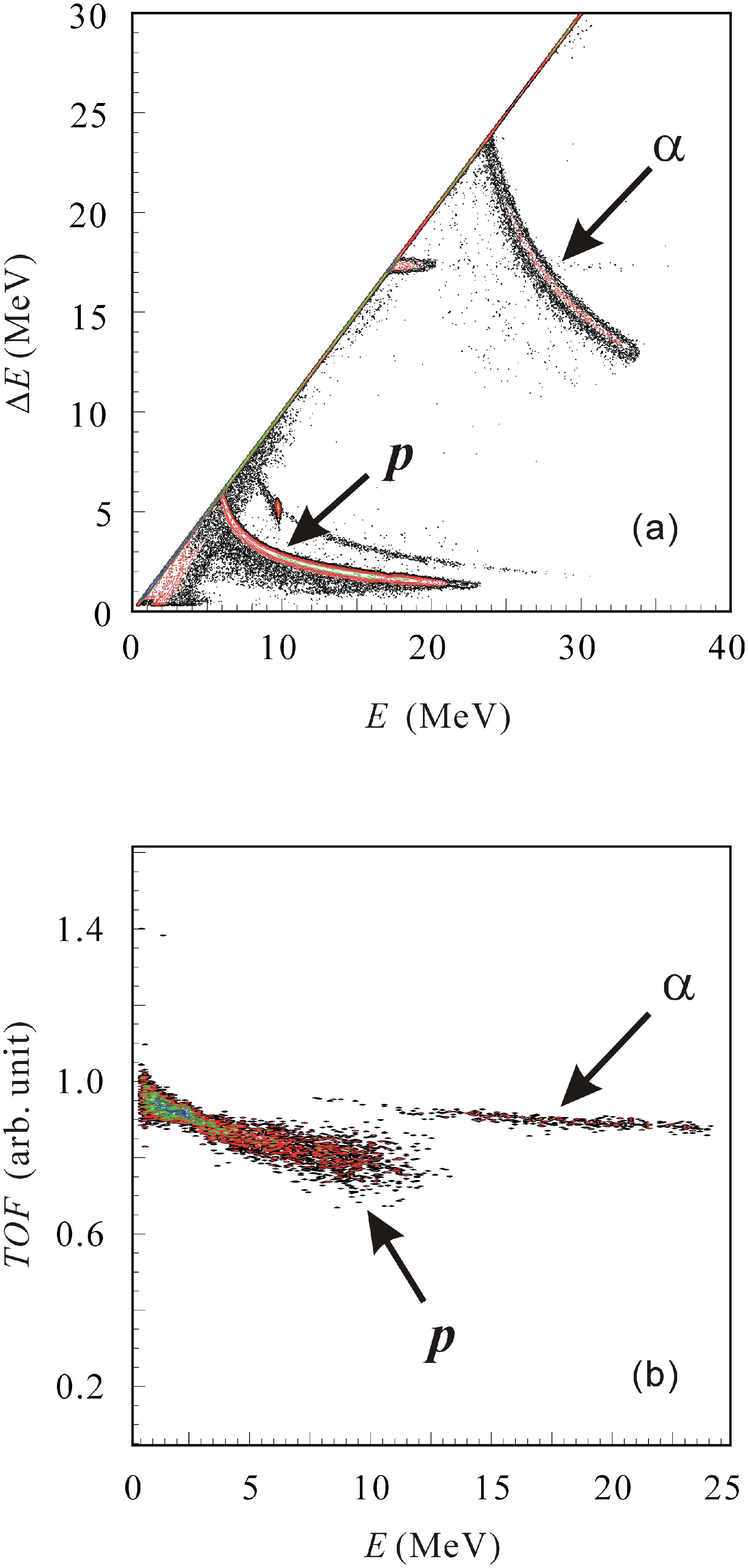}}
\caption{\label{fig3_pid} (Color online) Particle identification by (a) $\Delta E$-$E$ method, and (b) $TOF$-$E$ method. $\Delta E$ and $E$ signals are
measured by the silicon telescopes, and $TOF$ is the time-of-flight between PPACb and $\Delta$E. See text for details.}
\end{figure}

The recoiled light particles were detected with three $\Delta$E-E silicon telescopes at laboratory angles of
$\theta_\mathrm{Si}^{lab}$$\approx$0$^\circ$ (hereafter referred to as ``Set 1"), +14$^\circ$ (``Set 2") and -14$^\circ$ (``Set 3") with
respect to the beam line, respectively. These silicon detectors were produced by the Micron Semiconductor Inc.~\cite{micron}.
Each telescope subtended an opening angle of about 10$^\circ$ with a solid angle of about 27 mSr in the laboratory frame. In the $c.m.$ frame
for elastic scattering, the averaged scattering angles of the telescopes correspond to $\theta_{c.m.}$$\approx$175.2$^\circ$ (Set 1),
152.2$^\circ$ (Set 2) and 150.5$^\circ$ (Set 3), respectively. $\Delta$E is the position sensitive double-sided-strip (16$\times$16 strips,
3~mm width of each strip) detector, which measured the energy, position and timing signals of the light particles. The pad E detectors (1.5 mm
thick) measured their residual energies. This allowed for the clear identification of recoiled particles as shown in Fig.~\ref{fig3_pid}.
The high-energy particles penetrating through $\Delta$E can be identified by the $\Delta E$-$E$ method (see Fig.~\ref{fig3_pid}(a)); the
low-energy particles fully stopped in $\Delta$E can be identified by the $TOF$-$E$ method (see Fig.~\ref{fig3_pid}(b)), where $TOF$ is the
time-of-flight between PPACb and $\Delta$E. In this work, the energy calibration for the Si detectors was carried out by using secondary proton
beams produced with CRIB and a standard triple-$\alpha$ source.

Experimental data with a C target (13.5 mg/cm$^{2}$) was also acquired in a separate run to evaluate the contributions from the reactions of
$^{21}$Na with C nuclei. The yield ratio of these two proton spectra [with (CH$_{2})_{n}$ and C targets] was normalized by the number of beam
particles and by the target thickness per unit beam energy loss in the corresponding targets~\cite{he_al,he}.

\section{Results}
\label{sec:result}
\begin{figure}[!h]
\resizebox{8.3cm}{!}{\includegraphics{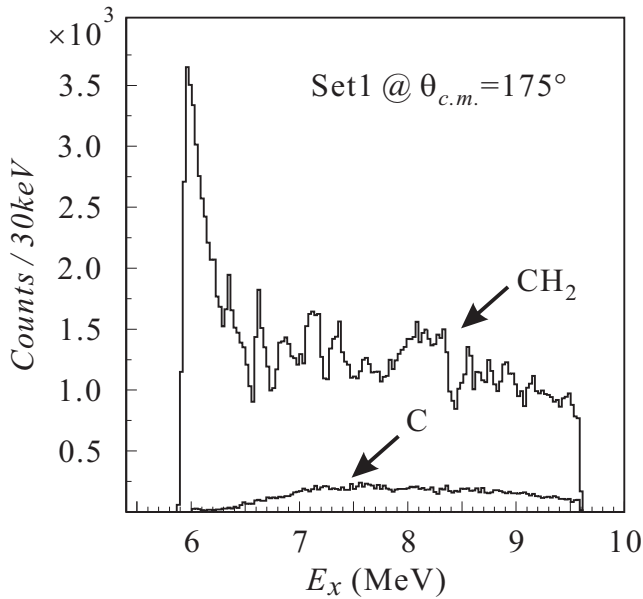}}
\caption{\label{fig4_psd1} Reconstructed proton spectrum for the $^{21}$Na+$p$ elastic scattering taken at
$\theta_{c.m.}$$\approx$175.2${^\circ}$ (Set 1). The abscissa is scaled to the excitation energy in $^{22}$Mg. The normalized carbon background
spectrum is also shown for comparison.}
\end{figure}

For inverse kinematics, the center-of-mass energy $E_{c.m.}$ of the $^{21}$Na+$p$ system is related to the energy $E_p$ of the recoiling protons
detected at a laboratory angle $\theta_{lab}$ by~\cite{he_al}
\begin{eqnarray}
E_{c.m.}=\frac{A_p + A_t}{4 A_p \mathrm{cos}^2 \theta_{lab}} E_p
\label{eq:one},
\end{eqnarray}
where $A_p$ and $A_t$ are the mass numbers of the projectile and target nuclei; this equation is valid only for an elastic scattering case.
In practice, $E_{p}$ was converted to $E_{c.m.}$ by assuming the elastic scattering kinetics and considering the energy loss of particles in
the target. A sample proton spectrum from the (CH$_{2}$)$_{n}$ target runs obtained at the scattering of $\theta_{c.m.}$$\approx$175.2$^\circ$
(Set 1) is shown in Fig.~\ref{fig4_psd1}.
The proton spectrum with a 13.5 mg/cm$^{2}$ C
target (C spectrum) is also shown for comparison. The C spectra can be described using smooth curves with respect to the energy in
all three telescopes. These C spectra were normalized to the corresponding proton spectra with the (CH$_{2}$)$_{n}$ target using the number
of beam particles and the number of C atoms per unit energy loss of the beam. The normalized yield in a C spectrum was about 1/6 of that in the
(CH$_{2}$)$_{n}$ spectrum at maximum (see Fig.~\ref{fig4_psd1}).

The laboratory differential cross section ($d\sigma$/$d\Omega$) for $^{21}$Na+$p$ scattering with energy $E_p$ and angle $\theta_{lab}$ is
deduced~\cite{he_al} through the proton spectrum by
\begin{eqnarray}
\frac{d\sigma}{d\Omega_{lab}}(E_p,\theta_{lab})=\frac{N}{I_0 N_s \Delta\Omega_{lab}},
\label{eq:two}
\end{eqnarray}
where $N$ is the number of detected protons, {\it i.e.}, at energy interval of $E_p \rightarrow E_p+ \Delta E$ and scattering angle of $\theta_{lab}$,
which are measured by a Si telescope covering a solid angle $\Delta\Omega_{lab}$. $I_0$ is the total number of $^{21}$Na beam particles that bombarded
the (CH$_2)_n$ target, and is considered to be constant in the whole energy region. $N_s$ is the number of H atoms per unit area per energy bin
in the target (${dx/dE}$)~\cite{zie85}. The transformation of the laboratory differential cross sections to the $c.m.$ frame is given by
\begin{eqnarray}
\frac{d\sigma}{d\Omega_{c.m.}}(E_{c.m.},\theta_{c.m.})=\frac{1}{4 \mathrm{cos} \theta_{lab}} \frac{d\sigma}{d\Omega_{lab}}(E_{p},\theta_{lab}).
\label{eq:three}
\end{eqnarray}

\begin{figure}[!h]
\resizebox{8.5cm}{!}{\includegraphics{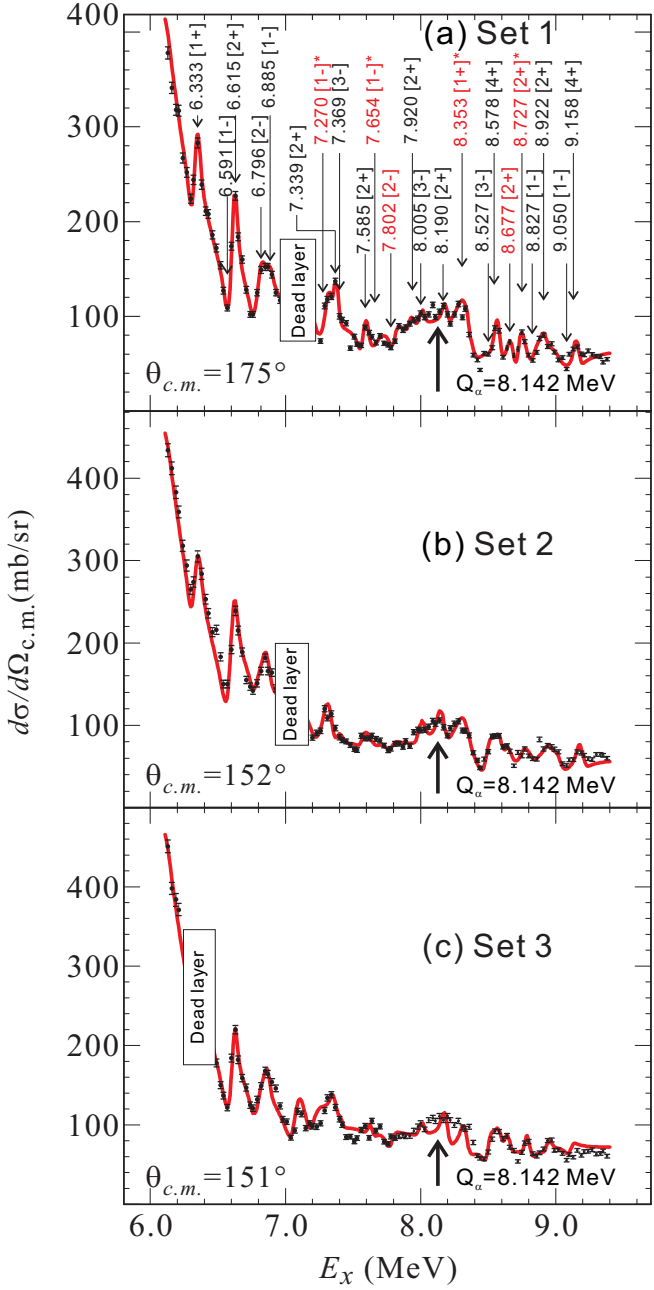}}
\caption{\label{fig5_fits} (Color online) The $c.m.$ differential cross sections for the resonant elastic scattering of $^{21}$Na+$p$ measured by
three sets of telescopes at different angles. The most probable $R$-matrix fitting results are shown. The data within the dead-layer region
(between $\Delta$E and E detectors) were removed. See text for details.}
\end{figure}

Fig.~\ref{fig5_fits} shows the {$c.m.$} differential cross sections for the resonant elastic scattering of $^{21}$Na+$p$ measured at angles of
$\theta_{c.m.}$$\approx$175.2${^\circ}$ (Set 1), 152.2${^\circ}$ (Set 2) and 150.5${^\circ}$ (Set 3), respectively. The dead-layer region shown
in Set 1\&2 is different from that in Set 3, simply because the thickness of $\Delta$E1\&2 (300 $\mu$m) is different from that of $\Delta$E3
(65 $\mu$m). The fitting in the dead-layer region (between $\Delta$ E and E detectors) are not reliable and removed from the figure. The abscissa is scaled by the excitation energies in $^{22}$Mg, which are calculated by $E_x$=$E_{c.m.}$+$Q_p$. As such, a value of
$Q_p$=5.504 MeV is adopted based on the updated masses of $^{21}$Na and $^{22}$Mg~\cite{mukherjee,wang}. Here, the energy resolution of $E_{c.m.}$
was determined by the resolution of the silicon detection system, the angular resolution of the scattering angle, as well as the energy width of
the secondary beam and the particle straggling in the target material. Thereinto, the detector energy resolution dominates the total energy resolution
of $E_{c.m.}$ in three telescopes. Based on a Monte-Carlo simulation, the overall energy resolution (FWHM) of $E_{c.m.}$ in Set 1 was estimated to be
$\sim$30 keV, while those in Set 2\&3 were about 30--70 keV (over $E_{c.m.}$=0.5--4 MeV) because of the larger scattering angle resulting in the larger
kinematics shifts. The error of the deduced cross-section is estimated to be 6\%, which mainly arises from the statistical error of the proton yields
and that of the target thickness. The deduced excitation energies in $^{22}$Mg indicated on Fig.~\ref{fig5_fits} are calculated by $E_x$=$E_R$+$Q_p$,
with resonance energy $E_{R}$ determined by the $R$-matrix analysis as discussed below. The present excitation energies agree with those adopted by
Matic {\it et al.} within the uncertainties (see discussion below).

The center-of-mass ($c.m.$) differential cross-section data have been analyzed by a multichannel $R$-matrix~\cite{lan58} code MULTI~\cite{nelson}. The
overall $R$-matrix fits are shown in Fig.~\ref{fig5_fits}. A channel radius of $R_n$=1.35(1+21$^\frac{1}{3}$) fm~\cite{gorres,chen} was adopted in the
calculation. All possible $R$-matrix attempts were restricted to $\ell$$\le$4, since resonances of higher $\ell$ transfer are invisible within the
present resolution. Here, it is worth mentioning that the experimental data at Set 2\&3 also support the $J^\pi$ assignments made for the Set 1,
although following figures shown below are the $R$-matrix fits on the Set 1 data. The parameters for the resonances in $^{22}$Mg deduced from the
present $R$-matrix analysis (Set 1) are summarized in Table~\ref{table1_fit}. The excitation energies and spin-parities deduced from this work are
compared to the previous ones in Tables~\ref{table2_below}\&\ref{table3_above}. In the following $R$-matrix fitting figures, the (red) solid lines
are the best fits with the parameters listed in Table~\ref{table1_fit}. The details of the $R$-matrix analysis will be discussed in the following subsections.

\begin{table}
\caption{\label{table1_fit} Resonant parameters used in the present $R$-matrix analysis. The errors of $E_x$ ($^{22}$Mg) and $\Gamma_{p}$ were estimated to be 12 keV and 18 keV respectively based on the standard deviation of the corresponding values in three sets.}
\begin{ruledtabular}
\begin{tabular}{ccccc}
 $E_x$ ($^{22}$Mg)      & $J^\pi$            & $s$           &  ${\ell}$        & $\Gamma_{p}$ (keV)      \\
\hline
6.333 & 1$^{+}$ & 1 & 0 & 16 \\
6.591 & 1$^{-}$ & 2 & 1 & 36 \\
6.615 & 2$^{+}$ & 2 & 0 & 10 \\
6.796 & 2$^{-}$ & 1 & 1 & 62 \\
6.885 & 1$^{-}$ & 2 & 3 & 2 \\
7.270 & 1$^{-}$ & 2 & 1 & 82 \\
7.339 & 2$^{+}$ & 2 & 2 & 18 \\
7.369 & 3$^{-}$ & 2 & 3 & 7 \\
7.585 & 2$^{+}$ & 2 & 0 & 16 \\
7.654 & 1$^{-}$ & 2 & 1 & 114 \\
7.802 & 2$^{-}$ & 1 & 1 & 19 \\
7.920 & 2$^{+}$ & 2 & 0 & 3 \\
8.005 & 3$^{-}$ & 2 & 3 & 1 \\
8.190 & 2$^{+}$ & 2 & 2 & 5 \\
8.353 & 1$^{+}$ & 1 & 2 & 97 \\
8.527 & 3$^{-}$ & 2 & 1 & 3 \\
8.578 & 4$^{+}$ & 2 & 2 & 5 \\
8.677 & 2$^{+}$ & 2 & 2 & 7 \\
8.727 & 2$^{+}$ & 2 & 0 & 12 \\
8.827 & 1$^{-}$ & 2 & 1 & 57 \\
8.922 & 2$^{+}$ & 2 & 2 & 4 \\
9.050 & 1$^{-}$ & 2 & 1 & 105 \\
9.158 & 4$^{+}$ & 2 & 2 & 2 \\
\end{tabular}
\end{ruledtabular}
\end{table}

\subsubsection{\label{sec:level1} Levels below the $\alpha$ threshold }
\begin{table*}
\caption{\label{table2_below} Excitation energies, spin-parities of levels below the $\alpha$ threshold in $^{22}$Mg.}
\begin{ruledtabular}
\begin{tabular}{cccccccc}
 Present & Matic {\it et al.}  & Caggiano {\it et al.} & Chen {\it et al.} & Chae {\it et al.} & Berg {\it et al.} & Ruiz {\it et al.} & He {\it et al.}\\
 $^{21}$Na+$p$ & ($p$,$t$)~\cite{matic}  & ($^{3}$He,$^{6}$He)~\cite{caggiano} & ($^{16}$O,$^{6}$He)~\cite{chen} & ($p$,$t$)~\cite{chae}  & ($^{4}$He,$^{6}$He)~\cite{berg} & $^{21}$Na+$p$~\cite{ruiz} & $^{21}$Na+$p$~\cite{he}\\
 \hline
6.333 1$^{+}$   & 6.306  (3$^{+}$)  & 6.329 (4$^{+}$) &   -             &  -              & -      & 6.333 1$^{+}$           & -   \\
6.591 1$^{-}$   & 6.578  (1$^{-}$)  & -               &   -             &  -              & -      & 6.591 1$^{-}$           & -   \\
6.615 2$^{+}$   & 6.602  (2$^{+}$)  & 6.616           & 6.606           &  -              & 6.606  & 6.615 2$^{+}$           & 6.61 2$^{+}$ \\
6.796 2$^{-}$   & 6.7688 (0$^{+}$)  & 6.771 (3$^{-}$) & 6.767 (3$^{-}$) &  -              & 6.766  & 6.796 (1$^{-}$,2$^{-}$) & 6.81 (1$^{+}$,2$^{+}$) \\
6.885 (1$^{-}$) & 6.8760 (1$^{-}$)  & 6.878           & 6.889 (3$^{-}$) &  -              & -      & 6.885                   & 6.93 (2$^{+}$,3$^{-}$) \\
7.270 (1$^{-}$) & 7.2183 (0$^{+}$)  & 7.206 (0$^{+}$) & 7.169 (0$^{+}$) &  -              & 7.216  & -                       & 7.27 (2$^{+}$,1$^{+}$) \\
7.339 (2$^{+}$) & 7.338  (2$^{+}$)  & -               &  -              & -               & -      & -                       & -\\
7.369 (3$^{-}$) & 7.389  (3$^{-}$)  & 7.373           & 7.402           &  -              & -      & -                       & 7.42 (1,2$^{+}$) \\
7.585 (2$^{+}$) & 7.5995 (2$^{+}$)  & 7.606           &  -              & 7.614           & -      & -                       & 7.59 (1$^{+}$,2$^{+}$) \\
7.654 (1$^{-}$) & -                 & -               & 7.674           &  -              & -      & -                       & -\\
7.802 (2$^{-}$) & 7.741  (4$^{+}$)  & 7.757           & 7.784           &  -              & -      & -                       & 7.82 (2$^{-}$) \\
7.920 (3$^{-}$) & 7.921             & -               & 7.964           & 7.967 (2$^{+}$) & 7.938  & -                       & 7.98 (2$^{+}$) \\
8.005 (3$^{-}$) & 8.007  (3$^{-}$)  & 7.986           & -               &  -              & -      & -                       & -              \\
\end{tabular}
\end{ruledtabular}
\end{table*}

\begin{figure}
\resizebox{8.2cm}{!}{\includegraphics{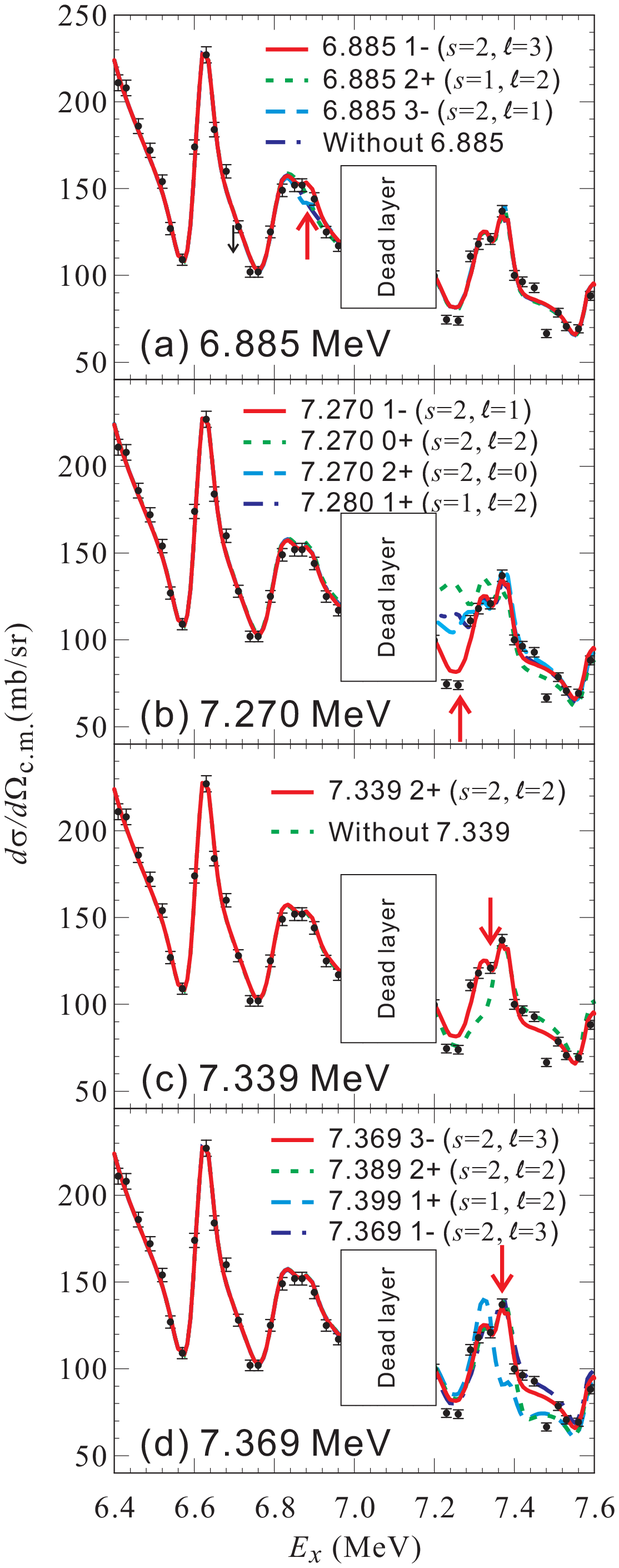}}
\caption{\label{fig6_below} (Color online) $R$-matrix fits for states below the $\alpha$ threshold in $^{22}$Mg.}
\end{figure}

\begin{figure}
\resizebox{8.2cm}{!}{\includegraphics{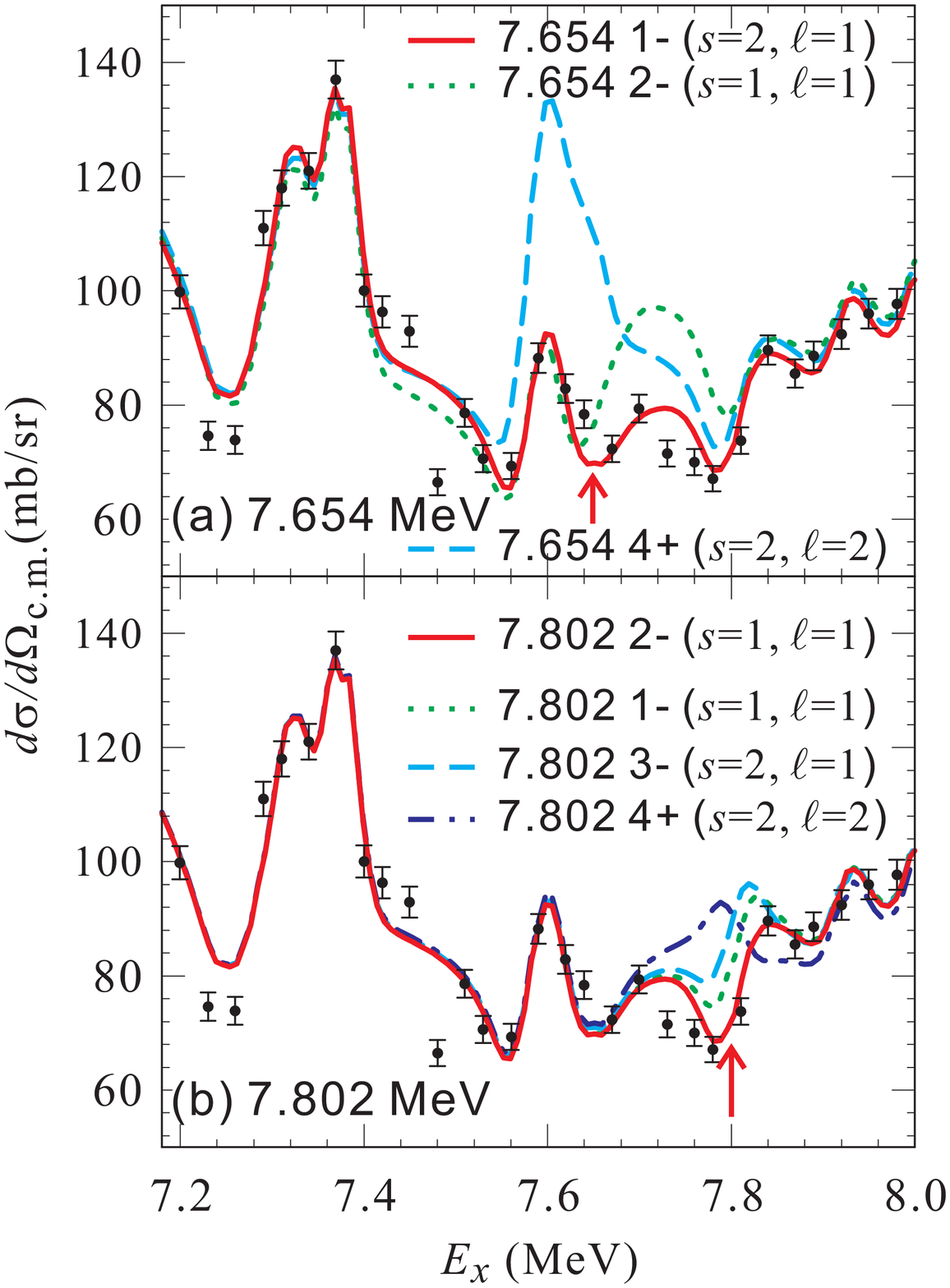}}
\caption{\label{fig7_below2} (Color online) $R$-matrix fits for states below the $\alpha$ threshold in $^{22}$Mg.}
\end{figure}

Four states observed at 6.333, 6.591, 6.615, 6.796 MeV were well studied before~\cite{ruiz}, and had been assigned as 1$^{+}$, 1$^{-}$, 2$^{+}$, 2$^{-}$,
respectively. As shown in Fig.~\ref{fig5_fits}, the resonant shape of these states has been successfully reproduced by using the previous parameters
determined in Ref.~\cite{ruiz}, and hence the previous $J^{\pi}$ assignments are confirmed. Such agreement provides confidence in the present analysis.

The observed 6.885 MeV state is closest to the 6.876 MeV state in Ref.~\cite{matic} and the 6.885 MeV state in Ref.~\cite{ruiz}.
Ruiz {\it et al.}~\cite{ruiz} regarded it as a very weak state and excluded it from their $R$-matrix fitting. The present $R$-matrix fits for this state
are shown in Fig.~\ref{fig6_below}(a). This weak-populated state ($\Gamma_{p}$$\approx$2 keV) can be fitted reasonably by 1$^{-}$, 2$^{+}$ and 3$^{-}$,
but with 1$^-$ being the most preferred assignment suggested by Matic {\it et al.}.

Matic {\it et al.} observed four states at 7.027, 7.045, 7.060 and 7.079 MeV. In the present experiment, these states are located
over the dead-layer region of Set 1\&2 in which the data are not reliable for the $R$-matrix analysis. In Set 3, the energy resolution achieved is not
able to resolve these four states. Therefore, these states are excluded from the present $R$-matrix analysis.

The 7.270 MeV state was tentatively assigned as $J^{\pi}$=(0$^{+}$, 1$^{+}$, 2$^{+}$) before~\cite{matic,he}. By varying the channel-spins, $\ell$ values
and proton widths for these three assignments, we found none of these assignments can reproduce the experimental data well. The present $R$-matrix analysis
supports a 1$^{-}$ assignment for this state (see Fig.~\ref{fig6_below}(b)).

We have confirmed the existence of the 7.339 MeV state first identified by Matic {\it et al.} Fig.~\ref{fig6_below}(c) shows the contrast of fittings
with and without this state. Thus, the tentative 2$^{+}$ assignment by Matic {\it et al.} is confirmed here.

\begin{figure}
\resizebox{8.0cm}{!}{\includegraphics{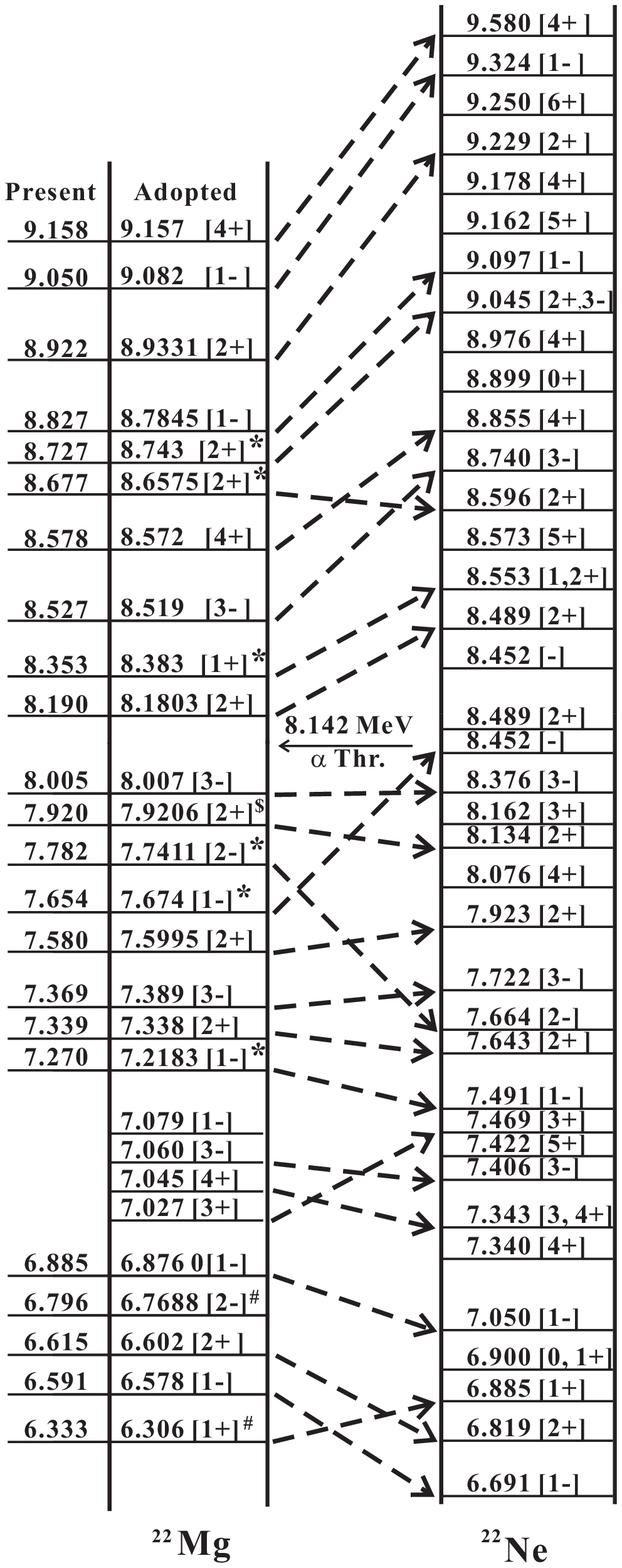}}
\caption{\label{fig8_level} Mirror assignments for the $^{22}$Mg levels above the proton threshold.
The adopted values (energies and $J^\pi$s) are mainly taken from Ref.~\cite{matic}. The new $J^\pi$ values assigned in this work
are marked by an $\ast$, while those assigned in Ref.~\cite{ruiz} and Ref.~\cite{chae} are marked by the \# and \$ symbols, respectively.
The structure data for the mirror $^{22}$Ne are adopted from Ref.~\cite{end90}.}
\end{figure}

The observed 7.369 MeV state is closest to the Matic {\em et al} 7.389 MeV (3$^{-}$) state.
It can be assigned as $J^{\pi}$=(3$^{-}$, 2$^{+}$), where 3$^{-}$ is preferred as shown in Fig.~\ref{fig6_below}(d). In addition, there
is no 2$^{+}$ state around this region in the mirror $^{22}$Ne (see Fig.~\ref{fig8_level}), and hence $J^{\pi}$=3$^{-}$ is assigned to this state.

Previously there had been five states observed in the excitation energy range of 7.5--8.1 MeV, at energies of 7.601, 7.674, 7.742, 7.921 and
8.005 MeV~\cite{matic}. Our data can be fitted with the corresponding excitation energies of 7.585, 7.654, 7.802, 7.920 and 8.005 MeV, respectively.
It is found that three resonances
at 7.585, 7.920 and 8.005 MeV can be fitted with 2$^+$, 2$^+$ and 3$^-$, the same assignments as suggested by Matic {\it et al.}, while the other two resonances
at 7.654 and 7.802 MeV can be preferentially fitted by 1$^-$ (see Fig.~\ref{fig7_below2}(a)) and 2$^-$ (see Fig.~\ref{fig7_below2}(b)).

\subsubsection{\label{sec:level1} Levels above the $\alpha$ threshold }
\begin{table*}
\caption{\label{table3_above} Excitation energies, spin-parities of levels above the $\alpha$ threshold in $^{22}$Mg.}
\begin{ruledtabular}
\begin{tabular}{clclccl}
Present work    & Matic {\it et al.}     & Caggiano {\it et al.}               & Chen {\it et al.}               & Chae {\it et al.}     & Berg {\it et al.}               & He {\it et al.}         \\
                & ($p$,$t$)~\cite{matic} & ($^{3}$He,$^{6}$He)~\cite{caggiano} & ($^{16}$O,$^{6}$He)~\cite{chen} & ($p$,$t$)~\cite{chae} & ($^{4}$He,$^{6}$He)~\cite{berg} & $^{21}$Na+$p$~\cite{he} \\
\hline
8.190 (2$^{+}$) & 8.1803 (2$^{+}$)     & 8.229                               & 8.203                           & -                     & 8.197                           & 8.18 (1$^{+}$-3$^{+}$)  \\
8.353 (1$^{+}$) & 8.383  (2$^{+}$)     & 8.934                               & 8.396                           & -                     & 8.380                           & 8.31 (1$^{+}$-3$^{+}$)  \\
8.527 (3$^{-}$) & 8.519  (3$^{-}$)     & 8.487                               & 8.547 (2$^{+}$)                 & 8.495 (2$^{+}$)       & 8.512                           & 8.51 (3$^{-}$)          \\
8.578 (4$^{+}$) & 8.572  (4$^{+}$)     & 8.598                               & -                               & -                     & -                               & -                       \\
8.677 (2$^{+}$) & 8.6575 (0$^{+}$)     & -                                   & 8.613 (3$^{-}$)                 & -                     & 8.644                           & 8.61 (2$^{+}$)          \\
8.727 (2$^{+}$) & 8.743  (4$^{+}$)     & -                                   & 8.754 (4$^{+}$)                 & -                     & -                               & -                       \\
8.827 (1$^{-}$) & 8.7845 (1$^{-}$)     & 8.789                               & -                               & -                     & 8.771                           & -                       \\
8.922 (2$^{+}$) & 8.9331 (2$^{+}$)     & -                                   & 8.925 (3$^{-}$)                 & -                     & 8.921                           & -                       \\
9.050 (1$^{-}$) & 9.082  (1$^{-}$)     & -                                   & 9.066                           & -                     & 9.029                           & -                       \\
9.158 (4$^{+}$) & 9.157  (4$^{+}$)     & -                                   & 9.172                           & -                     & 9.154                           & -                       \\
\end{tabular}
\end{ruledtabular}
\end{table*}

The $^{18}$Ne($\alpha$,$p$)$^{21}$Na reaction rate is determined by the $^{22}$Mg levels above the $\alpha$ threshold. The excitation energies of
these levels were very well studied before (see Ref.~\cite{matic} and references therein), but their spin-parities were still poorly known. In the
present experiment, ten resonances above the $\alpha$ threshold were observed and analyzed by the $R$-matrix code. We have experimentally confirmed
the $J^{\pi}$ values tentatively assigned by Matic {\it et al.} for seven states at 8.180, 8.519, 8.572, 8.785, 8.933, 9.082 and 9.157 MeV, and
assigned new  $J^{\pi}$ values for three states at 8.383, 8.658 and 8.743 MeV. The resonant parameters for calculating the $^{18}$Ne($\alpha$,$p$)$^{21}$Na
rate are summarized in Table~\ref{table3_above}. The present $J^{\pi}$ assignments will be discussed in detail below.

The observed 8.190 MeV state corresponds to the 2$^{+}$ state observed at 8.180 MeV by Matic {\it et al.} The $J^{\pi}$=(1$^{+}$--3$^{+}$) assignments
were suggested to the 8.18 MeV state by the $R$-matrix analysis of the previous $^{21}$Na+$p$ data~\cite{he}. Here, this state is still able to be
fitted by $J^{\pi}$=(1$^{+}$--3$^{+}$). We simply adopt the 2$^{+}$ assignment suggested by Matic {\it et al.} Anyway its contribution to the
total rate is negligible (see Table~\ref{table4_par}).

\begin{figure}
\resizebox{8.65cm}{!}{\includegraphics{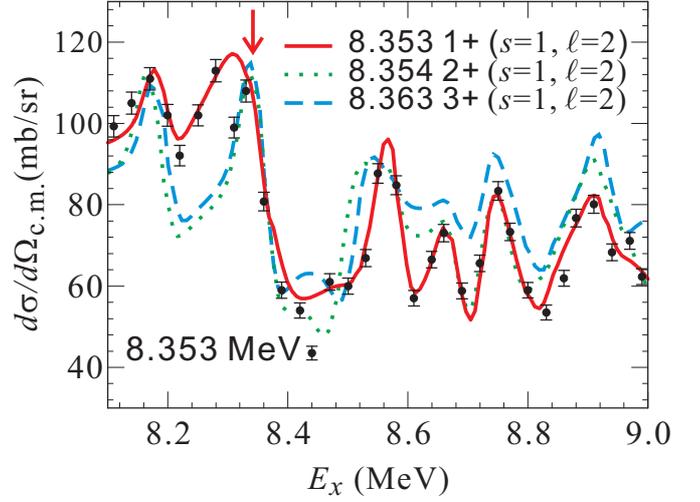}}
\caption{\label{fig9_8353} (Color online) $R$-matrix fits for the 8.353 MeV state.}
\end{figure}

The observed 8.353 MeV state was assigned as $J^{\pi}$=(1$^{+}$--3$^{+}$) in the previous $^{21}$Na+$p$ experiment~\cite{he}, where 1$^{+}$ was
suggested to be the most probable assignment. That assignment was only tentative due to the poor statistics. This state is close to the 8.383 MeV state
observed by Matic {\it et al.} who suggested a 2$^{+}$ assignment by referring to the mirror state in $^{22}$Ne. In the present $R$-matrix fit, 1$^{+}$
is the best candidate as shown in Fig.~\ref{fig9_8353}. Furthermore, this state was only weakly populated in the transfer-reaction
experiments~\cite{matic,chen,berg} which preferentially populated the natural-parity states in $^{22}$Mg. This again supports our unnatural-parity
1$^{+}$ assignment for this state. Therefore, this state does not contribute to the $^{18}$Ne($\alpha$,$p$)$^{21}$Na reaction rate.

\begin{figure*}
\resizebox{17cm}{!}{\includegraphics{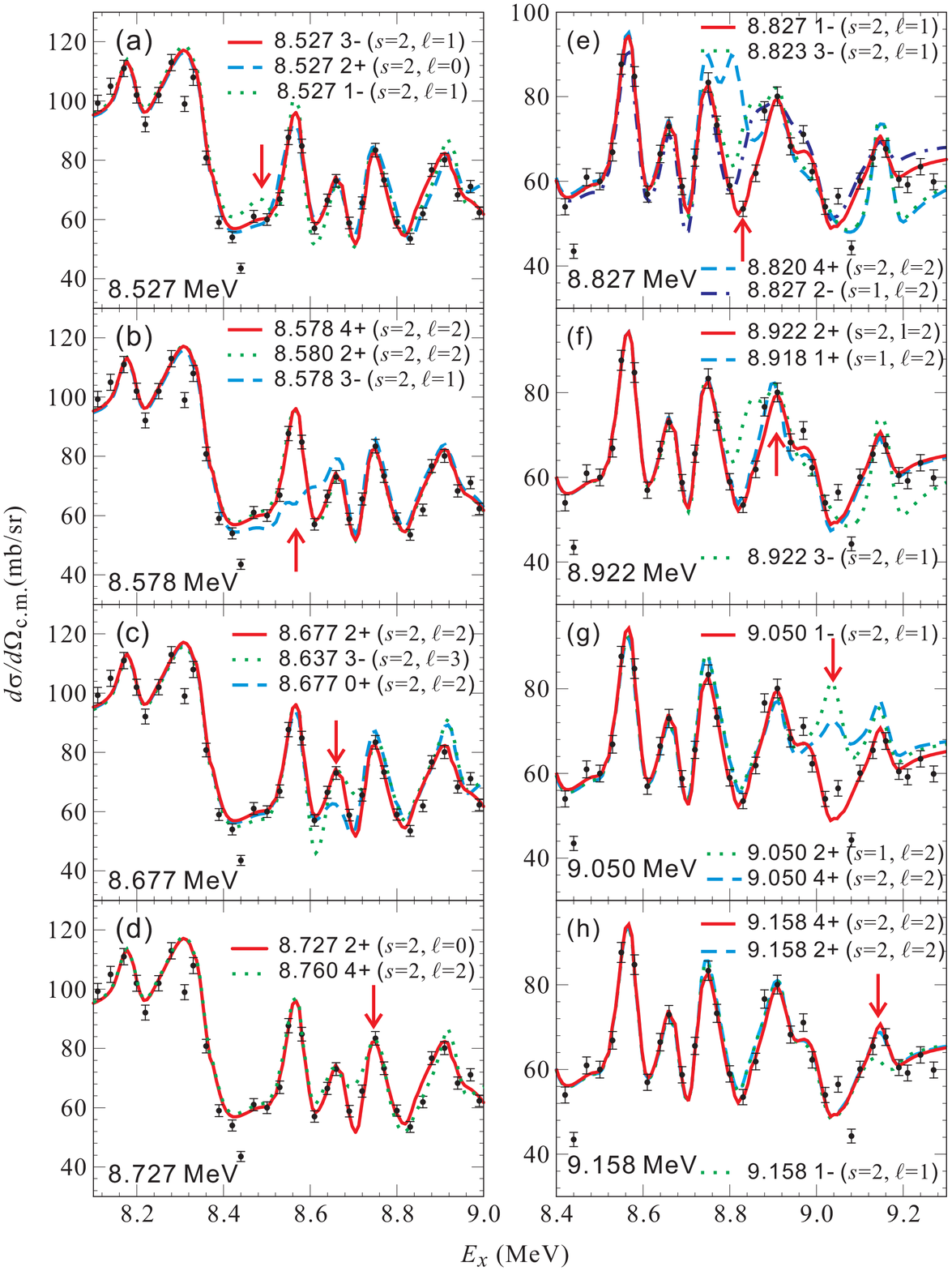}}
\caption{\label{fig10_above} (Color online) $R$-matrix fits for states above the $\alpha$ threshold.}
\end{figure*}

The observed 8.527 MeV state is close to the 3$^{-}$ state at 8.519 MeV by Matic {\it et al.} and at 8.51 MeV by He {\it et al.}~\cite{he}.
It is also close to the 2$^{+}$ state observed at 8.547 MeV by Chen {\it et al.} and at 8.495 MeV by Chae {\it et al.} In this work, both 3$^{-}$ and
2$^{+}$ can fit the experimental data as shown in Fig.~\ref{fig10_above}(a). As discussed in section I, the mirror 2$^{+}$ assignment made by Chen {\it et al.} might
be questionable; Chae {\it et al.} could not well resolve the triplet at 8.459, 8.578 and 8.667 MeV, and hence their 2$^{+}$ assignment is questionable
as well. Especially, there is no 2$^{+}$ state in the mirror $^{22}$Ne around this region. Here, we assign the 8.527 MeV state as $J^{\pi}$=3$^{-}$.

The observed 8.578 MeV state is closest to the 8.572 MeV state of Matic {\it et al.} in which it was assigned as 4$^{+}$ based on the shell model
calculation. As shown in Fig.~\ref{fig10_above}(b), both 4$^{+}$ and 2$^{+}$ can fit our data very well. As such, our data support the previous
4$^{+}$ assignment.

The observed 8.677 MeV state corresponds to the 8.658 MeV state of Matic {\it et al.}, which was assigned as a $J^{\pi}$=0$^{+}$ based on the shell
model calculation. However, such a prediction is questionable because of the high level-density at such a high excitation energy region.
Matic {\it et al.} regarded this state as the 8.613 MeV state observed by Chen {\it et al.} who assumed a 3$^{-}$ by simply shifting the energy of
the mirror 8.741 MeV state in $^{22}$Ne by about 130 keV. In addition, a 2$^{+}$ was tentatively assigned to the 8.61 MeV state in the previous
low-statistics experiment~\cite{he}. As shown in Fig.~\ref{fig10_above}(c), the present experiment strongly prefers the 2$^{+}$ assignment.

The observed 8.727 MeV state is regarded as the 8.743 MeV state of Matic {\it et al.} and the 8.754 MeV state of Chen {\it et al.} Both of them gave
a 4$^{+}$ assignment, and considered it the mirror 8.976 MeV state in $^{22}$Ne. The present $R$-matrix analysis, however, suggests a 2$^{+}$ for
this state (see Fig.~\ref{fig10_above}(d)), the mirror of 9.045 MeV state in $^{22}$Ne (see Fig.~\ref{fig8_level}).

The observed 8.827 MeV state is close to the 8.785 MeV state of Matic~{\it et al.} who gave a tentatively 1$^{-}$ assignment to this state. We have
tried all possible spin-parities, and verified very well the 1$^{-}$ assignment as shown in Fig.~\ref{fig10_above}(e).

The spin-parity for the observed 8.922 MeV state was simply assumed to be 2$^{+}$ by Matic {\it et al.} and 3$^{-}$ by Chen {\it et al.}, respectively.
In this work, both 1$^{+}$ and 2$^{+}$ can reproduce the data well as shown in Fig.~\ref{fig10_above}(f), where fittings with three possible $J^{\pi}$s
are shown. Thus, we suggest this state a 2$^{+}$.

The observed 9.050 and 9.158 MeV states correspond to the 9.082 and 9.157 MeV states of Matic {\it et al.} who gave the 1$^{-}$, 4$^{+}$ tentative
assignments, respectively. The previous tentative 1$^{-}$ assignment for the 9.050 MeV state is confirmed as shown in Fig.~\ref{fig10_above}(g).
For the 9.158 MeV state, both 2$^{+}$ and 4$^{+}$ can fit the experimental data as shown in Fig.~\ref{fig10_above}(h). Based on the discussion
made by Matic {\it et al.}, 4$^{+}$ is thus adopted for the 9.158 MeV state.

For the states assigned the same $J^{\pi}$ values as in Ref.~\cite{matic}, we have adopted the corresponding mirror assignments for the analogue states
in $^{22}$Mg and $^{22}$Ne suggested by Matic {\it et al.}; while for those assigned new $J^{\pi}$ values, we made the new mirror assignments accordingly.
The present mirror assignments are shown in Fig.~\ref{fig8_level}.

\section{Astrophysical implications}
\label{sec:implications}
The thermonuclear $^{18}$Ne($\alpha$,$p$)$^{21}$Na rate, when calculated via a full numerical integration of the energy-dependent cross section,
deviates by less than 10\%~\cite{matic2} from the rate calculated using a narrow resonance formalism~\cite{chen,matic}. Such deviation is
negligible compared to the total uncertainty (estimated below) of the present rate. In this work, the following narrow-resonance formalism has been
utilized for the rate calculations,
\begin{widetext}
\begin{eqnarray}
\label{ratefactor}
N_{A}\langle \sigma v \rangle_\mathrm{res}=1.54 \times 10^{11} (\mu T_9)^{-3/2} \sum_i (\omega\gamma)_i \times\mathrm{exp} \left (-\frac{11.605E_R^i}{T_9} \right) [\mathrm{cm^3 s^{-1} mol^{-1}}],
\end{eqnarray}
\end{widetext}
where $\mu$ is the reduced mass in units of amu, $E_R^i$ and $(\omega\gamma)_i$ (both in units of MeV) the energy and strength of individual resonance,
$T_{9}$ the temperature in units of 10$^9$ K ({\it i.e.}, GK). With the condition of $\Gamma_{\alpha}\ll\Gamma_{p}\approx\Gamma_\mathrm{tot}$ the resonant
strength is calculated by
\begin{eqnarray}
(\omega\gamma)_i=\omega \frac{\Gamma_{\alpha}\Gamma_{p}}{\Gamma_\mathrm{tot}}\simeq(2J_i+1)\Gamma_{\alpha},
\end{eqnarray}
where $J_i$ is the spin of the resonance in the compound nucleus $^{22}$Mg. The $\alpha$-particle partial width is calculated by
\begin{eqnarray}
\label{width}
\Gamma_{\alpha}=\frac{3\hbar^{2}}{\mu R^{2}_{n}} C^{2}S_{\alpha}\times P_{\ell}(E_R).
\end{eqnarray}
Here, the isospin Clebsch-Gordan coefficient $C$ for the $^{18}$Ne+$\alpha$ system is 1. Our rate calculation depends critically on the spectroscopic
factors $S_{\alpha}$ of the resonances, which are not known experimentally. Similar to the method used by Matic {\it et al.}, the corresponding
$\alpha$-spectroscopic factors are adopted from the mirror $^{22}$Ne whenever available; while for the states without known $S_{\alpha}$ values, the
average $S_{\alpha}$ values [{\it i.e.}, $S_{\alpha}$(0$^{+}$)=0.11, $S_{\alpha}$(1$^{-}$)=0.11, $S_{\alpha}$(2$^{+}$)=0.11, $S_{\alpha}$(3$^{-}$)=0.05,
and $S_{\alpha}$(4$^{+}$)=0.03] adopted by Matic {\it et al.} are utilized in the present calculations. Here, the errors of these adopted factors
are estimated as the standard deviation of the $S_{\alpha}$ factors for the states with same $J^{\pi}$: they are $\sigma(0^{+}$)=0.18, $\sigma(1^{-}$)=0.09,
$\sigma(2^{+}$)=0.19, $\sigma(3^{-}$)=0.02, and $\sigma(4^{+}$)=0.03, respectively. The Coulomb penetration factor $P_{\ell}$ on resonance is given by
\begin{eqnarray}
P_{\ell}(E_R)=\frac{kR_n}{[F_{\ell}(E)^2+G_{\ell}(E)^2]_{\vert {E_R; R_n}}}
\label{penetration},
\end{eqnarray}
where $k$=$\sqrt{2\mu E_R}$/$\hbar$ is the wave number; $F_{\ell}$ and $G_{\ell}$ are the regular and irregular Coulomb functions, respectively.

\begin{table*}
\caption{\label{table4_par} The resonant parameters utilized for the $^{18}$Ne($\alpha$,$p$)$^{21}$Na rate calculation. The energies ($E_x$ and $E_R$)
are exactly those adopted in Ref.~\cite{matic}.}
\begin{ruledtabular}
\begin{tabular}{ccccccc}
 $E_x$($^{22}$Mg) (MeV)  & $E_R$ (MeV)     & $J^{\pi}$  & $S_{\alpha}$ &  $\Gamma_{\alpha}$ (eV) & $\omega\gamma$ (eV)     \\
\hline
8.1812	&	0.039	&  $2^{+}\footnotemark[1]$  &	2.80$\times$10$^{-01}$  &	1.70$\times$10$^{-65}$  &	 8.53$\times$10$^{-65}\footnotemark[4]$ \\
8.385	&	0.243	&  $1^{+}\footnotemark[1]$  &	0                       &	0                       &	 0$\footnotemark[3]$                    \\
8.5193	&	0.377	&  $3^{-}\footnotemark[1]$  &	4.00$\times$10$^{-03}$  &	7.00$\times$10$^{-15}$  &	 4.87$\times$10$^{-14}\footnotemark[5]$ \\
8.574	&	0.432	&  $4^{+}\footnotemark[1]$  &	6.00$\times$10$^{-02}$  &	3.60$\times$10$^{-13}$  &	 3.26$\times$10$^{-12}\footnotemark[4]$ \\
8.6572	&	0.515	&  $2^{+}\footnotemark[1]$  &	3.20$\times$10$^{-01}$  &	2.10$\times$10$^{-08}$  &	 1.03$\times$10$^{-07}\footnotemark[3]$ \\
8.743	&	0.601	&  $2^{+}\footnotemark[1]$  &	1.10$\times$10$^{-01}$  &	2.70$\times$10$^{-07}$  &	 1.34$\times$10$^{-06}\footnotemark[3]$ \\
8.7832	&	0.642	&  $1^{-}\footnotemark[1]$  &	1.10$\times$10$^{-01}$  &	4.00$\times$10$^{-06}$  &	 1.21$\times$10$^{-05}\footnotemark[4]$ \\
8.9318	&	0.790	&  $2^{+}\footnotemark[1]$  &	1.10$\times$10$^{-01}$  &	8.30$\times$10$^{-05}$  &	 4.13$\times$10$^{-04}\footnotemark[4]$ \\
9.08	&	0.938	&  $1^{-}\footnotemark[1]$  &	1.10$\times$10$^{-01}$  &	7.70$\times$10$^{-03}$  &	 2.31$\times$10$^{-02}\footnotemark[4]$ \\
9.157	&	1.015	&  $4^{+}\footnotemark[1]$  &	7.80$\times$10$^{-02}$  &	9.70$\times$10$^{-05}$  &	 8.70$\times$10$^{-04}\footnotemark[4]$ \\
9.318	&	1.176	&  $2^{+}\footnotemark[2]$  &	1.10$\times$10$^{-01}$  &	9.90$\times$10$^{-02}$  &	 4.97$\times$10$^{-01}\footnotemark[4]$ \\
9.482	&	1.342	&  $3^{-}\footnotemark[2]$  &	1.50$\times$10$^{-02}$  &	1.80$\times$10$^{-02}$  &	 1.25$\times$10$^{-01}\footnotemark[4]$ \\
9.542	&	1.401	&  $1^{-}\footnotemark[2]$  &	1.10$\times$10$^{-01}$  &	4.40$\times$10$^{+00}$  &	 1.31$\times$10$^{+01}\footnotemark[4]$ \\
9.709	&	1.565	&  $0^{+}\footnotemark[2]$  &	1.50$\times$10$^{-01}$  &	5.20$\times$10$^{+01}$  &	 5.18$\times$10$^{+01}\footnotemark[4]$ \\
9.7516	&	1.610	&  $2^{+}\footnotemark[2]$  &	6.20$\times$10$^{-02}$  &	1.61$\times$10$^{+01}$  &	 4.82$\times$10$^{+01}\footnotemark[4]$ \\
9.86	&	1.718	&  $0^{+}\footnotemark[2]$  &	1.90$\times$10$^{-02}$  &	2.10$\times$10$^{+01}$  &	 2.07$\times$10$^{+01}\footnotemark[4]$ \\
10.085	&	1.944	&  $2^{+}\footnotemark[2]$  &	5.00$\times$10$^{-02}$  &	4.50$\times$10$^{+01}$  &	 2.25$\times$10$^{+02}\footnotemark[4]$ \\
10.2715	&	2.130	&  $2^{+}\footnotemark[2]$  &	1.10$\times$10$^{-01}$  &	2.62$\times$10$^{+02}$  &	 1.31$\times$10$^{+03}\footnotemark[4]$ \\
10.429	&	2.287	&  $4^{+}\footnotemark[2]$  &	3.00$\times$10$^{-02}$  &	5.43$\times$10$^{+00}$  &	 4.89$\times$10$^{+01}\footnotemark[4]$ \\
10.651	&	2.509	&  $3^{-}\footnotemark[2]$  &	5.00$\times$10$^{-02}$  &	1.60$\times$10$^{+02}$  &	 1.12$\times$10$^{+03}\footnotemark[4]$ \\
10.768	&	2.626	&  $2^{+}\footnotemark[2]$  &	1.10$\times$10$^{-01}$  &	2.30$\times$10$^{+03}$  &	 1.16$\times$10$^{+04}\footnotemark[4]$ \\
10.873	&	2.731	&  $0^{+}\footnotemark[2]$  &	1.10$\times$10$^{-01}$  &	1.19$\times$10$^{+04}$  &	 1.19$\times$10$^{+04}\footnotemark[4]$ \\
11.001	&	2.859	&  $4^{+}\footnotemark[2]$  &	3.00$\times$10$^{-02}$  &	6.45$\times$10$^{+01}$  &	 5.81$\times$10$^{+02}\footnotemark[4]$ \\
11.315	&	3.173	&  $4^{+}\footnotemark[2]$  &	3.00$\times$10$^{-02}$  &	2.00$\times$10$^{+02}$  &	 1.83$\times$10$^{+03}\footnotemark[4]$ \\
11.499	&	3.357	&  $2^{+}\footnotemark[2]$  &	1.10$\times$10$^{-01}$  &	1.70$\times$10$^{+04}$  &	 8.64$\times$10$^{+04}\footnotemark[4]$ \\
11.595	&	3.453	&  $4^{+}\footnotemark[2]$  &	3.00$\times$10$^{-02}$  &	4.08$\times$10$^{+02}$  &	 3.67$\times$10$^{+03}\footnotemark[4]$ \\
11.747	&	3.607	&  $0^{+}\footnotemark[2]$  &	1.10$\times$10$^{-01}$  &	7.10$\times$10$^{+04}$  &	 7.13$\times$10$^{+04}\footnotemark[4]$ \\
11.914	&	3.772	&  $2^{+}\footnotemark[2]$  &	1.10$\times$10$^{-01}$  &	3.53$\times$10$^{+04}$  &	 1.77$\times$10$^{+05}\footnotemark[4]$ \\
12.003	&	3.861	&  $1^{-}\footnotemark[2]$  &	2.10$\times$10$^{-01}$  &	1.40$\times$10$^{+05}$  &	 4.31$\times$10$^{+05}\footnotemark[4]$ \\
12.185	&	4.050	&  $3^{-}\footnotemark[2]$  &	1.80$\times$10$^{-01}$  &	3.70$\times$10$^{+04}$  &	 2.60$\times$10$^{+05}\footnotemark[4]$ \\
12.474	&	4.332	&  $2^{+}\footnotemark[2]$  &	1.10$\times$10$^{-01}$  &	7.80$\times$10$^{+04}$  &	 3.89$\times$10$^{+05}\footnotemark[4]$ \\
12.665	&	4.523	&  $3^{-}\footnotemark[2]$  &	1.20$\times$10$^{-01}$  &	4.90$\times$10$^{+04}$  &	 3.45$\times$10$^{+05}\footnotemark[4]$ \\
13.01	&	4.865	&  $0^{+}\footnotemark[2]$  &	1.10$\times$10$^{-01}$  &	2.20$\times$10$^{+05}$  &	 2.16$\times$10$^{+05}\footnotemark[4]$

\end{tabular}
\end{ruledtabular}
\footnotetext[1]{Experimentally determined spin-parities in this work.}
\footnotetext[2]{Spin-parities assumed in Ref.~\cite{matic2} as adopted in the present work.}
\footnotetext[3]{Recalculated $\Gamma_\alpha$ and $\omega\gamma$ values in this work.}
\footnotetext[4]{Resonant strengths in Ref.~\cite{matic2} as adopted in the present work.}
\footnotetext[5]{Resonant strength in Ref.~\cite{matic} as adopted in the present work.}

\end{table*}

\begin{figure}
\resizebox{8.1cm}{!}{\includegraphics{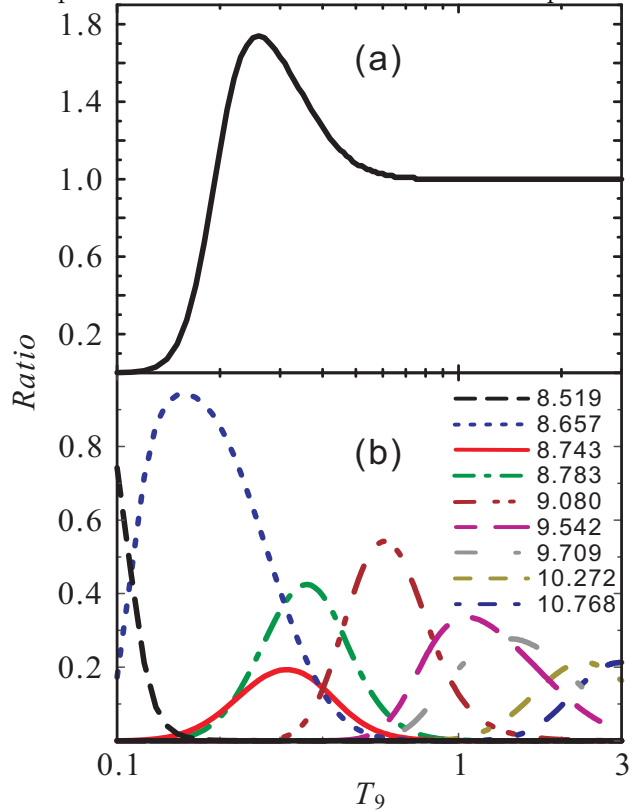}}
\caption{\label{fig11_rate} (Color online) (a) Ratio between the present $^{18}$Ne($\alpha$,$p$)$^{21}$Na rate and
that in Ref.~\cite{matic2}; (b) Contribution from those dominant resonances to the total rate.}
\end{figure}

The resonant parameters for the thermonuclear $^{18}$Ne($\alpha$,$p$)$^{21}$Na rate calculations are summarized in Table~\ref{table4_par}. In this
calculation, all resonant energies $E_R$ (and their errors) and most of the strengths $\omega\gamma$ are adopted from the work of
Matic {\it et al.}~\cite{matic} and Mohr \& Matic~\cite{matic2}. For those states assigned new $J^\pi$ values by this work, the strengths are
recalculated (see Table~\ref{table4_par}). Similar to the work of Mohr \& Matic, the calculated rate is taken as
$N_{A}$$\langle\sigma v\rangle$$_\mathrm{reference}$. The uncertainty of the calculated rate is mainly caused by the errors in the resonant strengths
$\omega\gamma$; the uncertainty arising from those of the resonant energies contributes less than 12$\%$ over 0.1--3 GK. This is verified by the
Monte-Carlo approach~\cite{RatesMC}. The two inverse measurements~\cite{salter,anl} gave similar rates which are about 3 times lower than the reference
rate at 0.8--2.7 GK~\cite{matic2}. As discussed in Ref.~\cite{matic2}, both calculations of the present work and the previous time-inverse measurement work are based on simple but reasonable arguments, if we assume the corresponding uncertainties do not exceed a factor of 2, there will be a relatively narrow overlap region between the lower
limit of the present $N_{A}$$\langle$$\sigma v$$\rangle$$_\mathrm{reference}$ and the upper limit of the reverse reaction data. The most realistic estimate from the overlap is located around $0.55\times N_{A}<\sigma~ v>_{reference}$, and is considered as the new recommended reaction rate:
$N_{A}$$\langle$$\sigma v$$\rangle$$_\mathrm{recommended}$ ~\cite{matic2}.
Note that the factor of 0.55 has been derived from the comparison of transfer data and reverse reaction data at energies between 1--3 MeV, {\it i.e.},
corresponding to temperatures above 1 GK where the Mohr \& Matic rate is practically identical to the present calculated one (see Fig.~\ref{fig11_rate}(a)).
Thus, the normalization factor of 0.55 is retained since the previous work. The calculated $N_{A}$$\langle$$\sigma v$$\rangle$$_\mathrm{recommended}$ rate is
summarized in Table~\ref{table5_rate}. As discussed by Mohr \& Matic, the realistic lower limit of the recommended rate can be taken from the Salter
{\it et al.} data (multiplied by a factor of 3 to take the ground-state branching into account), which is about three times lower than the reference
rate~\cite{salter}, and a realistic upper limit is the reference rate $N_{A}$$\langle$$\sigma v$$\rangle$$_\mathrm{reference}$.

\begin{table}
\caption{\label{table5_rate} Recommended $^{18}$Ne($\alpha$,$p$)$^{21}$Na reaction rate and the lower\&upper limits. All are in units of
cm$^{3}$ s$^{-1}$ mol$^{-1}$.}
\begin{ruledtabular}
\begin{tabular}{cccc}

$T_{9}$ & $N_{A}$$\langle$$\sigma v$$\rangle_\mathrm{recommended}$ & upper                     & lower  \\
\hline
0.1 	&	2.96$\times$10$^{-27}$ &	5.38$\times$10$^{-27}$ &	1.62$\times$10$^{-27}$  \\
0.2 	&	2.02$\times$10$^{-15}$ &	3.67$\times$10$^{-15}$ &	1.10$\times$10$^{-15}$  \\
0.3 	&	4.88$\times$10$^{-11}$ &	8.87$\times$10$^{-11}$ &	2.66$\times$10$^{-11}$  \\
0.4 	&	1.41$\times$10$^{-08}$ &	2.57$\times$10$^{-08}$ &	7.70$\times$10$^{-09}$  \\
0.5 	&	7.84$\times$10$^{-07}$ &	1.43$\times$10$^{-06}$ &	4.28$\times$10$^{-07}$  \\
0.6 	&	1.73$\times$10$^{-05}$ &	3.15$\times$10$^{-05}$ &	9.45$\times$10$^{-06}$  \\
0.7 	&	2.09$\times$10$^{-04}$ &	3.79$\times$10$^{-04}$ &	1.14$\times$10$^{-04}$  \\
0.8 	&	1.70$\times$10$^{-03}$ &	3.09$\times$10$^{-03}$ &	9.27$\times$10$^{-04}$  \\
0.9 	&	1.03$\times$10$^{-02}$ &	1.87$\times$10$^{-02}$ &	5.62$\times$10$^{-03}$  \\
1.0 	&	4.84$\times$10$^{-02}$ &	8.81$\times$10$^{-02}$ &	2.64$\times$10$^{-02}$  \\
1.1 	&	1.83$\times$10$^{-01}$ &	3.32$\times$10$^{-01}$ &	9.96$\times$10$^{-02}$  \\
1.2 	&	5.73$\times$10$^{-01}$ &	1.04$\times$10$^{+00}$ &	3.13$\times$10$^{-01}$  \\
1.3 	&	1.55$\times$10$^{+00}$ &	2.81$\times$10$^{+00}$ &	8.44$\times$10$^{-01}$  \\
1.4 	&	3.70$\times$10$^{+00}$ &	6.73$\times$10$^{+00}$ &	2.02$\times$10$^{+00}$  \\
1.5 	&	8.03$\times$10$^{+00}$ &	1.46$\times$10$^{+01}$ &	4.38$\times$10$^{+00}$  \\
1.6 	&	1.61$\times$10$^{+01}$ &	2.92$\times$10$^{+01}$ &	8.77$\times$10$^{+00}$  \\
1.7 	&	3.02$\times$10$^{+01}$ &	5.49$\times$10$^{+01}$ &	1.65$\times$10$^{+01}$  \\
1.8 	&	5.37$\times$10$^{+01}$ &	9.76$\times$10$^{+01}$ &	2.93$\times$10$^{+01}$  \\
1.9 	&	9.14$\times$10$^{+01}$ &	1.66$\times$10$^{+02}$ &	4.99$\times$10$^{+01}$  \\
2.0 	&	1.50$\times$10$^{+02}$ &	2.73$\times$10$^{+02}$ &	8.18$\times$10$^{+01}$  \\
2.1 	&	2.38$\times$10$^{+02}$ &	4.33$\times$10$^{+02}$ &	1.30$\times$10$^{+02}$  \\
2.2 	&	3.68$\times$10$^{+02}$ &	6.70$\times$10$^{+02}$ &	2.01$\times$10$^{+02}$  \\
2.3 	&	5.56$\times$10$^{+02}$ &	1.01$\times$10$^{+03}$ &	3.03$\times$10$^{+02}$  \\
2.4 	&	8.21$\times$10$^{+02}$ &	1.49$\times$10$^{+03}$ &	4.48$\times$10$^{+02}$  \\
2.5 	&	1.19$\times$10$^{+03}$ &	2.16$\times$10$^{+03}$ &	6.49$\times$10$^{+02}$  \\
2.6 	&	1.69$\times$10$^{+03}$ &	3.08$\times$10$^{+03}$ &	9.24$\times$10$^{+02}$  \\
2.7 	&	2.37$\times$10$^{+03}$ &	4.32$\times$10$^{+03}$ &	1.30$\times$10$^{+03}$  \\
2.8 	&	3.28$\times$10$^{+03}$ &	5.96$\times$10$^{+03}$ &	1.79$\times$10$^{+03}$  \\
2.9 	&	4.47$\times$10$^{+03}$ &	8.13$\times$10$^{+03}$ &	2.44$\times$10$^{+03}$  \\
3.0 	&	6.01$\times$10$^{+03}$ &	1.09$\times$10$^{+04}$ &	3.28$\times$10$^{+03}$
\end{tabular}
\end{ruledtabular}
\end{table}

The ratios between our recommended rate and the one recommended in Ref.~\cite{matic2} are shown in Fig.~\ref{fig11_rate}(a). It shows that the present
rate is much smaller below 0.13 GK. This is due to the unnatural-parity 1$^+$ newly assigned to the 8.385 MeV state which does not contribute to the rate
anymore. In addition, the present rate is about 1.7 times larger around 0.2 GK, because of our new $2^{+}$ assignments for the 8.657 and 8.743 MeV states.
Beyond 0.55 GK, the present rate is quite similar to the previously recommended one because the same resonant parameters for the high-lying states were used.
The contributions of those dominant resonances to the total rate are shown in Fig.~\ref{fig11_rate}(b). The resonance strength of the 8.743 MeV state
increases by more than two orders of magnitude as a result of the new spin-parity assignment, which significantly enhances its contribution compared with
the previous estimate of Matic {\it et al.} A comparison to other available rates was made in Ref.~\cite{he_prc_rc}, where the deduced rate is slightly
different from the present recommended one.

The recommended rate $N_{A}$$\langle$$\sigma v$$\rangle$$_\mathrm{recommended}$ (in units of cm$^3$s$^{-1}$mol$^{-1}$) can be well parameterized
(within 2\% error in 0.1--10 GK) by the following expression, {\it e.g.}, Eq.~(16) in Ref.~\cite{rau00}:
\begin{widetext}
\begin{eqnarray}
N_A\langle\sigma v\rangle &=&\mathrm{exp}(-2201.74-401.30T_9^{-1}+5496.69T_9^{-1/3}-2970.73T_9^{1/3}+73.64T_9-2.135T_9^{5/3}+2374.71\ln{T_9}) \nonumber \\
                          &+&\mathrm{exp}(5493.46-410.12T_9^{-1}+2889.36T_9^{-1/3}-9603.61T_9^{1/3}+1544.42T_9-286.244T_9^{5/3}+3066.95\ln{T_9}) \nonumber \\
                          &+&\mathrm{exp}(13426.5-119.86T_9^{-1}+8781.47T_9^{-1/3}-24789.9T_9^{1/3}+3149.91T_9-452.62T_9^{5/3}+8688.98\ln{T_9}) \nonumber
\label{eq:fit}
\end{eqnarray}
\end{widetext}

The impact of our new $^{18}$Ne($\alpha$,$p$)$^{21}$Na rate was examined within the framework of one-zone XRB postprocessing calculations~\cite{he_prc_rc}.
Different XRB thermodynamic histories were employed, including the K04 ($T_\mathrm{peak}$=1.4 GK) and S01 ($T_\mathrm{peak}$=1.9 GK) models from
Refs.~\cite{par08,par09}. For each of these histories, separate postprocessing calculations were performed using the present $^{18}$Ne($\alpha$,$p$)$^{21}$Na
rate and previous ones~\cite{gorres,chen,matic,he}; all other reaction rates in the network~\cite{par08} were left unchanged.

Our previous conclusion~\cite{he_prc_rc} about the energy generation rate affected by this $^{18}$Ne($\alpha$,$p$)$^{21}$Na rate still holds
with the present slightly changed rate. As concluded before, our new thermonuclear $^{18}$Ne($\alpha$,$p$)$^{21}$Na rate clearly affects predictions
from our models. For example, a striking difference in the nuclear energy generation rate at early times (between 0.3 and 0.4 s, or equivalently,
between 0.6 GK and 0.9 GK during the burst) has been seen when comparing XRB calculations using the present, Chae {\it et al.} and
G\"{o}rres {\it et al.} rates with the K04 model. Not only does the peak energy generation rate increase by a factor of 1.4--1.8 with the
present rate, but the profiles of the curves around the maxima are also rather different (see Fig. 4 in Ref.~\cite{he_prc_rc}).

\begin{figure}
\resizebox{8.0cm}{!}{\includegraphics{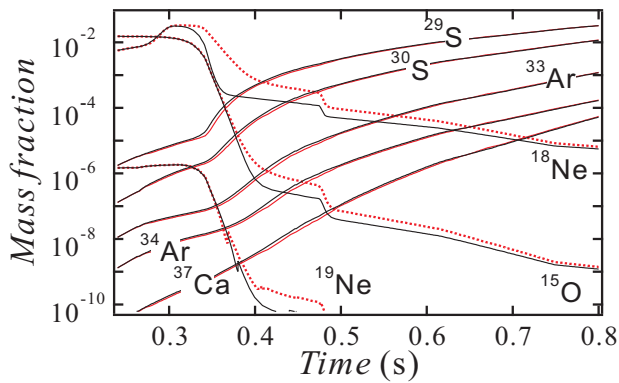}}
\caption{\label{fig12_abundance} (Color online) Abundances during one-zone XRB calculations using the K04 thermodynamic
history~\cite{par08}. Results using the present rate (black solid lines) and G\"{o}rres {\it et al.}~\cite{gorres} rate (red dotted lines) are indicated.}
\end{figure}

We note that for both the K04 and S01 models, rates from Refs.~\cite{gorres,chen,he} give lower peak nuclear energy generation rates than
that from Chae {\it et al.}, by about 10--30~\%. Furthermore, the rate of Matic {\it et al.} gives rather similar results to those using the present rate.
In particular, the calculated nuclear energy generation rates agree overall to about 5\%~\cite{he_prc_rc}. Therefore, we only discuss below the impact of
the G\"{o}rres {\it et al.} and present rates on the reaction flux in the K04 model. We note a change in the $^{18}$Ne($\alpha$,$p$)$^{21}$Na reaction flux
at the early times of the XRBs as shown in Fig.~\ref{fig12_abundance}. For example, at 0.35 s, this reaction flux increases by a factor of 2--3 with our new
rate. This contributes to the depletion of $^{15}$O and $^{18}$Ne at that time by a factor of 3--4 relative to abundances calculated using the G\"{o}rres
{\it et al.} rate. These species are effectively converted to higher mass ones. Note, however, that no significant changes ($>$5\%) to any final abundances
with mass fractions $X$$>$10$^{-3}$ were observed when comparing the calculations using the two rates (our recommended one and G\"{o}rres {\it et al.} one).

\begin{figure}[!htbl]
\resizebox{8.0cm}{!}{\includegraphics{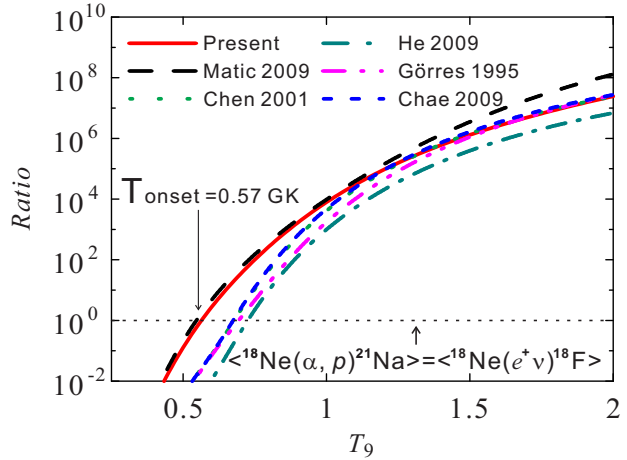}}
\caption{\label{fig13_decay} (Color online) Present rate and previous ones in unit of the $\beta$-decay rate of $^{18}$Ne($e^{+}$$v$)$^{18}$F.}
\end{figure}

Figure~\ref{fig13_decay} shows the ratio of the reaction rates of $^{18}$Ne($\alpha$,$p$)$^{21}$Na to the $\beta$-decay of
$^{18}$Ne($\beta^{+}$ $v$)$^{18}$F \cite{iliadis},
\begin{eqnarray}
\label{ratio}
R=\rho \frac{X_{\alpha}}{m_{\alpha}} N_{A}\langle \sigma v \rangle/(\mathrm{ln2}/\tau_{1/2}),
\end{eqnarray}
here, $\tau_{1/2}$ is the $\beta$-decay lifetime of $^{18}$Ne, $X_{\alpha}$ and $m_{\alpha}$ are the mass fraction and atomic mass of $\alpha$ particle, assuming a typical density 10$^6$~g/cm$^3$ and $\alpha$ mass fraction of 0.27 of XRBs.
It shows that present $^{18}$Ne($\alpha$,$p$)$^{21}$Na reaction dominates over the $\beta$-decay of $^{18}$Ne at an onset temperature of $T$$\approx$0.57 GK.
This critical temperature is noticeably lower than the temperature of $T$$\approx$0.68 GK with the rates from Refs.~\cite{gorres,chen,chae}, and hence
it implies that this reaction initiates the breakout earlier than previously thought. Note, the above numbers 0.57 and 0.68 were, by mistake, presented
as 0.47 and 0.60 in our previous publication~\cite{he_prc_rc}.

\section{Summary}
We have studied the resonant elastic-scattering of $^{21}$Na+$p$ using a radioactive ion beam of $^{21}$Na with the thick-target method. The $E_{c.m.}$
spectra were reconstructed from the inverse kinematics. In total, 23 resonances above the proton-threshold in $^{22}$Mg were observed. The relevant proton
resonant parameters have been determined by the $R$-matrix analysis of the center-of-mass differential cross-section data at different scattering angles.
The $^{18}$Ne($\alpha$,$p$)$^{21}$Na reaction rate is recalculated with the present parameters. A new recommended rate is given by combining the results
from different experimental techniques, and our new rate deviates considerably from the recent recommended rate of Mohr \& Matic below $\sim$0.55 GK.

The astrophysical impact of our new rate has been investigated through one-zone postprocessing x-ray burst calculations. Compared to previous rates in
Refs.~\cite{gorres,chen,chae}, the new rate increases the energy production rate by factors of 1.4--1.8 at early time (between 0.3--0.4 s, or equivalently,
between 0.6--0.9 GK during the burst) of the burst, and the $^{18}$Ne($\alpha$,$p$)$^{21}$Na reaction flux is also enhanced about two times at that time.
The breakout onset temperature for this reaction occurs at around 0.57 GK, lowered by 0.11 GK due to the increase of the reaction rate.

Despite the different $J^\pi$ values adopted in the present and Matic {\it et al.} $^{18}$Ne($\alpha$,$p$)$^{21}$Na rate calculations (and the
consequent differences in deduced thermonuclear rates), our models give very similar XRB nuclear energy generation rates. This suggests that $J^\pi$
values for relevant states in $^{22}$Mg are, for the moment, sufficiently well known for our models. Future measurements should primarily focus on
measuring other quantities of interest (such as spectroscopic factors, partial widths or the precise cross section data), which can further constrain this rate.

\begin{center}
\textbf{Acknowledgments}
\end{center}

We would like to thank the RIKEN and CNS staff for their friendly operation of the AVF cyclotron. This work was financially
supported by the National Natural Science Foundation of China (Nos. 11135005, 11021504), the Major State Basic Research
Development Program of China (2013CB834406), as well as the JSPS KAKENHI (No. 21340053, 25800125). AP was supported by the Spanish MICINN
(Nos. AYA2010-15685, EUI2009-04167), the E.U. FEDER funds as well as the ESF EUROCORES Program EuroGENESIS.

\end{document}